\documentclass[lettersize,journal]{IEEEtran}
\usepackage{amsmath,amsfonts}
\usepackage{algorithmic}
\usepackage{algorithm}
\usepackage{array}
\usepackage[caption=false,font=normalsize,labelfont=sf,textfont=sf]{subfig}
\usepackage{textcomp}
\usepackage{stfloats}
\usepackage{url}
\usepackage{verbatim}
\usepackage{hyperref}
\usepackage{graphicx}
\usepackage{xcolor}
\usepackage{cite}
\usepackage{booktabs} 
\usepackage{tabularx} 
\usepackage[T1]{fontenc}
\begin{document}

\title{Unpacking the process of watching a 360-degree video: How physiological responses and head rotation relate to affective and cognitive outcomes}

\author{Portia Wang, Yujie Tao, Monique Santoso, and Jeremy Bailenson
\thanks{Portia Wang, Yujie Tao, Monique Santoso, and Jeremy Bailenson are with Stanford University, Stanford, CA, 94305, USA (email:portiaw@stanford.edu; yjtao@stanford.edu; msant@stanford.edu; bailenson@stanford.edu).}}


\maketitle

\begin{abstract}
A defining feature of 360-degree video is that the perceptual experience is actively constructed. As viewers turn their head, shift their gaze, and encounter changing visuals, their perceptual experience is not predetermined, but rather emerges through ongoing explorations. Yet, it is unclear how the process of watching immersive videos unfold, and how it is related to psychological and physiological outcomes. We present a large-scale study (N=200) in which participants experienced one of 10 pieces of 360-degree videos in virtual reality. Results showed that greater head rotation was associated with greater valence, higher attention focus, longer length of subjective review and lower simulator sickness; greater pupil dilation was associated with lower valence, lower review ratings, and greater use of negative tone in written reviews. A mediation analysis revealed that participants who rotated their heads more experienced greater frame-to-frame pixel-level perceptual changes in their field of view, which in turn resulted in greater valence and lower simulator sickness. Moreover, participants tended to rotate their heads toward field of view regions with greater perceptual movement, while directing their eye gaze toward less perceptually dynamic areas within that view. We outline theoretical and practical contributions for designing 360-degree content and predicting experiential outcomes.
\end{abstract}


\section{Introduction}
Watching video content is a prevalent facet of everyday life. Videos are used to tell stories, share information, and establish common ground for social connection. At the same time, the technologies used to deliver content have greatly expanded in past years, ranging from mobile platforms such as TikTok to immersive ones such as virtual reality (VR) and multisensory domes like the Sphere in Las Vegas. As viewing technologies have diversified, so too have the forms of content people consume, varying in duration, format, and level of immersion, from brief 2D clips to longer-form 360-degree videos \cite{mabrook2019}.

Many scholars and practitioners have examined video viewing through 2D content \cite{goldstein2007, wu2018}, though it is unclear whether these insights will generalize to 360-degree content. Unlike 2D videos, which present viewers with a fixed viewing perspective, 360-degree videos allow viewers to actively shape what they see by changing their viewing direction. In VR, this process unfolds especially naturally, because movements of the head and body are directly reflected in the virtual environments. This affordance fundamentally changes the viewing experience. Rather than passively receiving the same visual information, viewers of 360-degree content construct their own perceptual experience. As a result, two individuals can watch the same video yet attend to drastically different parts of the scenes, encounter different visual cues, and ultimately form divergent impressions of the video. Indeed, differences in exploration patterns during brief 360-degree videos in VR are linked to variation in user experience and evaluation \cite{jun2020}.

\begin{figure*}[t]
 \centering 
 \includegraphics[width=0.92\textwidth,keepaspectratio]{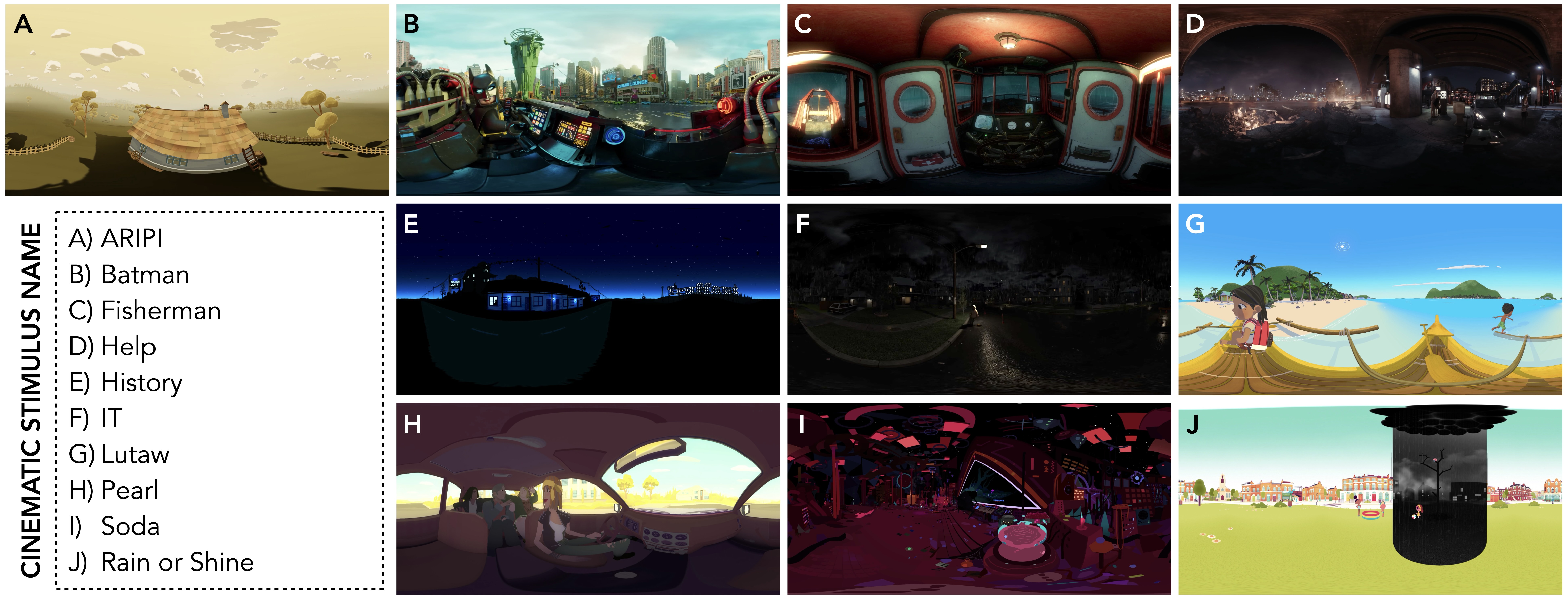}
 \caption{Screenshots from the Ten Sampled Immersive Videos. Fictional pieces are chosen to control for prior familiarity. To study narrative engagement, all videos had complete narrative arcs and are animated and computer-generated.}
 \label{fig: stimuli}
\end{figure*}

For this reason, understanding how people watch 360-degree videos presents both theoretical and practical challenges. From a theoretical perspective, 360-degree video raises new questions about topics such as attention, perception, and subjective experience when viewers are given greater control over what they see. From a practical perspective, creators need a better understanding of the viewing process to craft immersive content that remains engaging for audiences with diverse viewing tendencies. Yet, important gaps remain in our understanding of how individuals navigate, interpret, and psychologically respond to 360-degree video. This work presents a large-scale study of 200 participants, with each participant watching one of ten 360-degree videos in VR (see Figure \ref{fig: stimuli} for video screenshots). By collecting and analyzing head rotation, gaze direction, physiological signals of heart rate, heart rate variability, and pupil sizes, as well as self-report affective and cognitive outcomes, we make three contributions. 

First, we establish that moment-to-moment behaviors such as head rotation and pupil dilation are closely associated with the viewer’s affective and cognitive outcomes such as valence, simulator sickness, and tone of the written reviews. Our second set of contributions are methodological and descriptive; we propose measures for quantifying perceptual differences using frame differences and one’s head rotation and eye gaze data. We further establish a coding scheme for characterizing movement of the virtual camera and virtual elements in 360-degree videos. Using these measures, we describe the content characteristics of our examined videos, as well as participant’s viewing strategies as reflected through the perceptual differences in their field of view and foveal region. Our final contribution substantiates the empirical value of the proposed measures. For example, we demonstrate that perceptual difference mediates the relationship between head rotation and the outcomes of valence and simulator sickness and further that moment-to-moment content characteristics are associated with behavioral and physiological signals.

\section{Related works}
\subsection{Modeling user experiences during immersive experiences } 
Prior work suggests that behavioral signals can provide useful insights on how viewers experience immersive media. In VR, head rotation is often employed as a proxy for exploration, reflecting the degree to which a user actively samples their surroundings . Studies of 360-degree video have linked this form of exploration to subjective experience, although the direction of the relationship has varied across contexts. For example, increased head rotation has been associated with lower preferences for brief 20-second clips \cite{jun2020}, but with greater valence in longer 360-degree videos ranging from 30 seconds to 10 minutes \cite{li2017}. Likewise, there is also evidence that individuals who move their heads more often during immersive experiences are more likely to report lower levels of simulator sickness \cite{walker2010}. Rather than pointing to a simple relationship between exploration and experience, these findings suggest that head movement may depend on other factors such as content type and perceptual demand. 

A distinct body of research examined how immersion level and other technical characteristics shape user experience. Compared to less immersive formats such as 2D screens and still images, VR experiences tend to elicit stronger emotional responses \cite{bujic2023, kuhne2023}, greater presence, and more vivid memories \cite{kisker2021, schone2019, szita2018}. Others have shown that subjective outcomes such as perceived quality and cybersickness can be predicted from content and user factors, including resolution, rendering device, user interests, and prior VR familiarity \cite{anwar2020}.

One particularly relevant technical consideration in 360-degree videos is the use of virtual camera movement, such as smooth camera translation or instantly shifting camera locations through cuts \cite{kudra2019}. Although these techniques can guide attention and support narrative progression \cite{nielsen2016, rothe2019}, they can also increase spatial disorientation \cite{rahimi2020} and discomfort, as mismatches between a viewer’s visual perception of an immersive experience and their vestibular responses are a well-documented cause for simulator sickness \cite{fernandes2016, yao2014}. Taken together, these works suggest that behavioral responses and subjective responses are shaped not only by their own tendencies to engage with immersive experiences, but also by the perceptual and technical properties of the experience itself.

Researchers have also used physiological signals to infer psychological and cognitive states during immersive experiences. For example, Luong and Holz \cite{luong2022} designed VR experiences to elicit fear and frustration and found that pupillary, respiratory, and electrodermal responses differed across these states. More broadly, physiological measures such as heart rate and pupillometry have been used to build machine learning models for predicting affective responses in immersive environments \cite{bulagang2021, luong2022, xue2023}. Specifically, pupil dilation is a useful measure for cognitive load in immersive contexts \cite{eckert2021, lee2024, wei2025}, while heart rate variability is used to assess stress responses \cite{brouwer2011, ham2017}. That said, much of this work centered on short-form videos or experiences specifically designed to induce certain affective states, leaving open how these signals operate during longer-form 360-degree videos with meaningful storylines.

This distinction is important as viewers may engage with longer narrative experiences differently than brief videos. In 2D media contexts, short- and long-form videos differ in patterns of engagement and interaction: some studies suggest that short-form videos produce lower engagement than longer videos \cite{chen2024}, whereas others find that they draw engagement away from long-form content \cite{rajendran2024} and generate more views and likes but fewer comments than regular videos \cite{violot2024}. With much prior VR research focused on shorter-form experiences \cite{luong2022, bulagang2021}, our goal is not to directly contrast long- and short-form videos, but to provide a comprehensive account of how users watch longer-form narrative 360-degree videos.

This focus is especially meaningful for immersive media, as short-form VR experiences are largely designed to elicit monotonic perceptual or affective responses, rather than sustained narrative experiences. As a result, they provide limited opportunities to examine constructs such as narrative engagement or linguistic patterns in written reviews. By studying longer-form 360-degree videos with complete storylines, we examine behavioral and physiological responses in a context closer to narrative media consumption. Methodologically, longer videos produce more reliable measurement of physiological signals such as heart rate variability \cite{shaffer2020}, allowing us to uniquely characterize how viewers experience immersive narratives.

Formally, we build on prior research by first examining behavioral and physiological signals and characterizing how they jointly relate to affective and cognitive outcomes. Moreover, we moved beyond predicting downstream self-reports alone by investigating the perceptual processes (i.e., frame differences) that may underlie these relationships. Finally, by sampling ten 360-degree videos with complete storylines, ranging between 4 and 15 minutes, we extend prior work beyond short clips or stimuli to elicit emotions. This enabled us to begin to develop a vocabulary for describing how content characteristics, particularly those related to camera and character movements, may shape the experience of watching 360-degree videos.

\subsection{Predicting user behaviors during 360-degree videos}
Researchers have also examined how users engage with 360-degree videos, with many leveraging machine learning for predicting gaze and head movements. Many have sought to use saliency maps for understanding perceptual attention \cite{aladagli2017, xu2018, zhu2020}. Early work of Xu et al. \cite{xu2018} demonstrated the efficacy of predicting gaze point locations in 360-degree videos using saliency maps extracted at three visual scopes of the panoramic image, field of view and foveal region of a previous timestamp. Building on this, Zhu et al. \cite{zhu2020} adopted a similar approach and applied the use of saliency maps to predict head and gaze behaviors in 360-degree images. Researchers have also adopted other approaches to modeling attention in 360-degree videos, for example using only previous time series of head movement to predict the upcoming head movement trajectories \cite{lee2021}. Taken together, this research reflects the field’s long-standing interest in understanding gaze and head movement, as evidenced through the releases of large-scale datasets of user behaviors in looking at 360-degree images \cite{li2021} and watching one-to-two minute 360-degree videos \cite{agtzidis2019, jin2022}. 

The current work extends prior works in several ways. To start, while past literature have largely emphasized the predictive accuracy of user movement using black-box approaches, we focus on building a theoretical understanding of the relationship between head and gaze patterns and content characteristics, as well as downstream psychological outcomes. Moreover, while prior work has largely focused on short 360-degree clips or static images, we argue that using 360-degree videos with complete narrative arcs enhances external validity and enables a more comprehensive assessment of emotion, engagement, and user strain such as simulator sickness. This research bridges two approaches: analyses of user behaviors based on a single VR video \cite{bender2019, farmer2021} and those drawn from large-scale samples of 360-degree videos \cite{jin2022, xu2018}.

\subsection{Characterizing 360-degree videos}
Scholars have largely described 360-degree videos on a content level, rather than how they unfold for viewers over time. Early research developed conceptual vocabularies for 360-degree video by emphasizing its use of audio and montage, and capacity to support exploration across space and time \cite{palmer2020, elmezeny2018}. Related work proposed guidelines for how content \emph{should} be structured, arguing that creators must balance attentional guidance with viewer freedom and acclimation \cite{dooley2017}. That said, these works remain primarily descriptive and prescriptive, offering limited insight into how content characteristics are \emph{actually} instantiated from moment to moment.

This emphasis on content-level description is especially pronounced in work on immersive journalism. Prior studies have categorized 360-degree news by genre, point of view, journalistic presence, and autonomy \cite{paino2019, mabrook2019}, while a more recent work coded technological, narrative, and interactive dimensions such as user control of pace and perspective, as well as the use of text and music \cite{de2022}. Although these frameworks are useful for comparing types of immersive media, they operate at the level of the entire narrative pieces and are less suited for explaining how viewers encounter changing perceptual information during viewing. This limitation is important because outcomes such as attention, affect, and discomfort likely depend on what is available to the viewer at a given moment, not only on how a piece is overall characterized.

A smaller body of work has quantified visual properties of 360-degree videos more directly, but primarily for technical evaluation. For example, De Simone et al. \cite{de2019} measured scene complexity using edge detection applied to cubic maps extracted from 360-degree frames, and Afzal et al. \cite{afzal2017} analyzed motion vectors to explain bit rate variability. While these approaches demonstrate that video properties can be operationalized, whole-frame measures also do not fully reflect what a viewer actually sees in VR, where only a subset of the whole scene is visible to a view at a given moment. Moreover, not all motion is equally meaningful for experience, as movement tied to narrative elements and the virtual camera may matter more than background scene motion.

We build on these efforts by moving beyond coarse categorizations of 360-degree videos toward a more fine-grained, viewer-centered account of the perceptual experience. Instead of relying on broad content categories or whole-frame technical metrics, we quantify moment-to-moment changes in the viewer’s field of view and foveal region, and introduce a coding scheme that captures the movement of narrative elements and the virtual camera. In doing so, we shift the focus from content-level description to temporal measures of how narrative 360-degree videos are perceived and experienced.

\begin{figure*}[t]
 \centering 
 \includegraphics[width=0.88\textwidth,keepaspectratio]{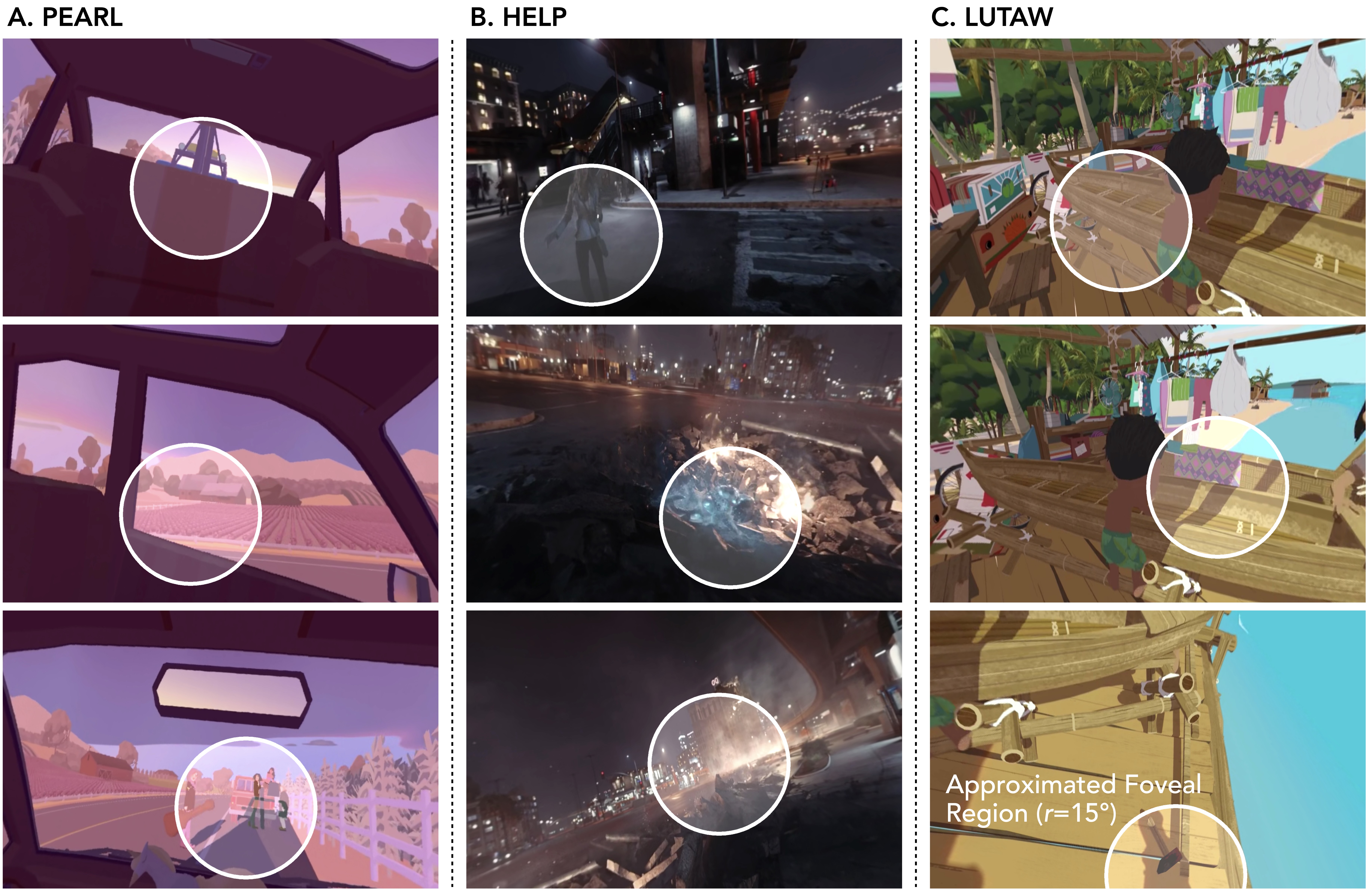}
 \caption{Sample User Gaze and Attention during Video Viewing. Each panel samples from participants who watched the same video and denotes the viewing perspectives of three users taken at the same timestamp of the video. The white circles denote the approximated foveal radius of 15 degrees \cite{patney2016}. The user’s perspective is reconstructed using the 360-degree video and the tracked head rotation and gaze behaviors. Notably, participants experienced the same video differently and used head and eye movements to direct their attention to different visual elements.}
 \label{fig: gaze examples}
\end{figure*}

\begin{figure*}[t]
 \centering 
 \includegraphics[width=0.88\textwidth,keepaspectratio]{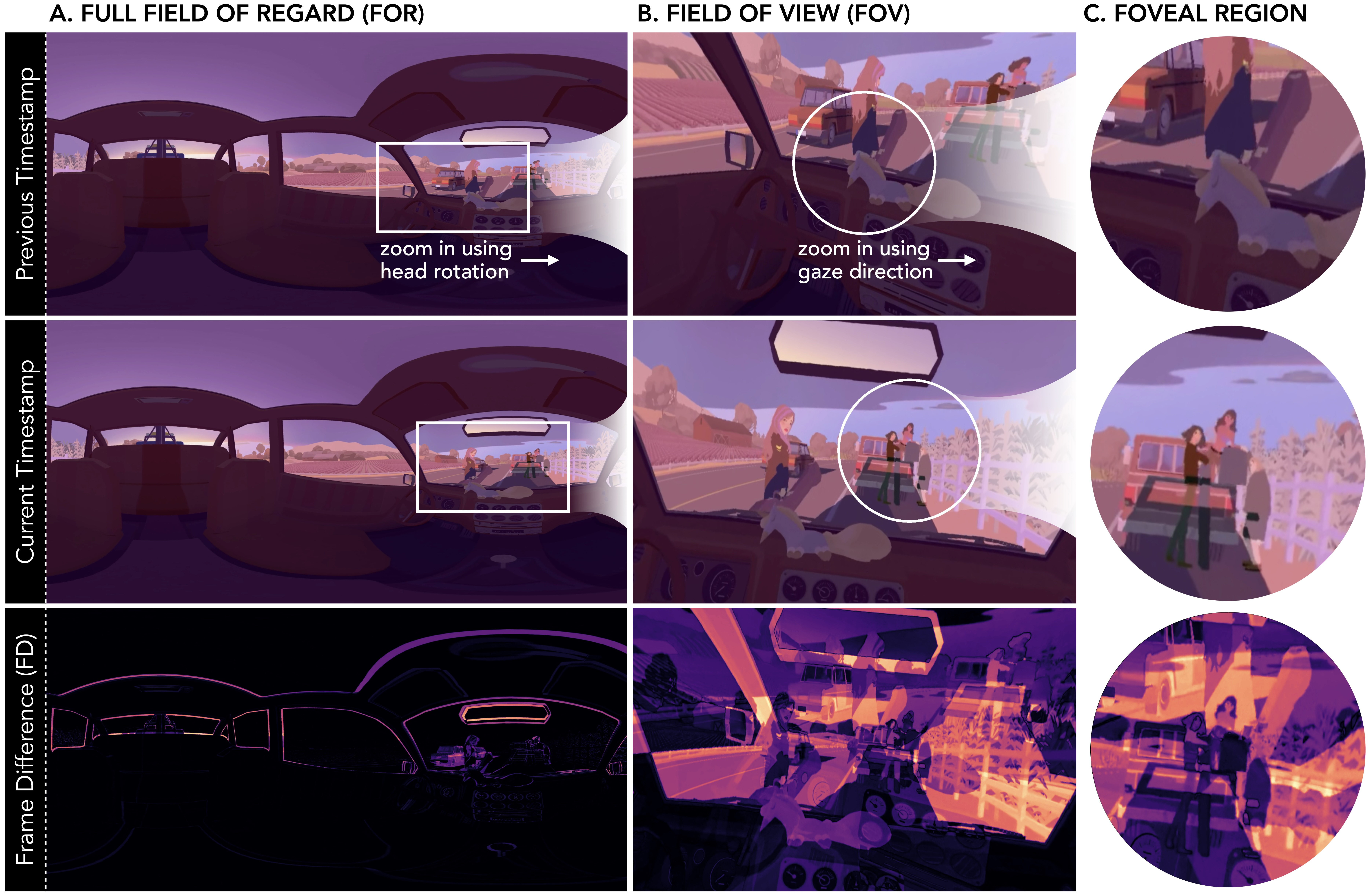}
 \caption{Sample Participant Frame Differences by Visual Scope. The figure shows data extracted from a participant viewing Pearl. Panels A–C show the three types of visual scope examined in our work. The top row represents regions extracted from a previous timestamp, the middle row denotes those from the current timestamp ($\Delta t$=0.5) and the bottom row visualizes per-pixel differences, averaged across color channels, before they are collapsed into a single value.}
 \label{fig: frame differences}
\end{figure*}

\section{Capturing perceptual experience and content characteristics}
\subsection{Measuring what users see using frame difference}
We used frame-differencing to quantify what individuals see while watching the 360-degree video (see Figure \ref{fig: gaze examples} for sample gaze patterns). Frame-differencing is largely used to assess the amount of movement and differences between consecutive frames to quantify movement \cite{singla2014}, and more specifically, human motion \cite{ramseyer2020}. At a high level, we first extract sequences of frames based on a predetermined extraction procedure for each viewer. Then, we calculated the absolute differences across each of the RGB color channels (i.e., red, green, blue) between pairs of consecutive frames. We finally averaged the matrix representing the absolute difference into a single value across all pixels that ranged between 0 and 255, where 0 represents no change, and 255 represents a total inversion of color (e.g., black to white). The values extracted across all pairs of consecutive frames are further averaged to derive a single measure per participant representing frame-differences. We downsampled consecutive frames at 2Hz to reduce computation time. Importantly, there are several ways for extracting and defining consecutive frames, and we elaborate on the two possible variations used in this work.

\subsubsection{Visual Scope} There are three ways for formulating our examined visual scope \cite{xu2018}: (1) the full field of regard (FOR), which consists of the entire 360-degree frame which the user could see through head rotation, (2) headset-driven field of view (FOV), and (3) a gaze-driven foveal region. Figure \ref{fig: frame differences} visualizes these three ways for segmenting an examined frame. In this way, we can quantify how much a participants’ perceived differences in their fovea and field of view may possibly differ from the average amount of “action” occurring in the entire 360-degree environment as a whole. More specifically, framing by the full FOR is achieved by extracting the raw frame of the 360-degree video. Framing by FOV is achieved by reconstructing the user’s viewing perspective based on their head orientation (i.e., pitch, yaw, roll), assuming a horizontal FOV of 94 degrees and a vertical FOV of 60 degrees that is consistent with our measurements of FOV using the Unity program. Framing by foveal region takes an additional step after reconstructing the user’s viewing perspective by first mapping the user’s gaze direction (combined across both eyes) in the viewing plane. The foveal region is then created by masking any areas of the FOV located more than 15 degrees from the center of the gaze. 15 degrees is chosen as the threshold following past literature on foveated rendering \cite{patney2016}.

\subsubsection{Attribution of Head Rotation and Stimuli Movement}The second variation decomposes frame differences into two components, namely changes caused by active head rotation (e.g., user looking around the scene), and those caused by changes in the stimuli (e.g., scene transition, camera cut). But first, frame differences between time $t-1$ and $t$ that do not make such a distinction and solely quantifies the differences in a user’s viewing perspective is defined as

\begin{equation}
\begin{aligned}
FD_{total}(t) = \frac{1}{3P} \sum_{c \in \{R,G,B\}} \sum_{i=1}^P & \left| F_{i,c}(S_t,R_t) \right. \\
& \left. - F_{i,c}(S_{t-1},R_{t-1}) \right|
\end{aligned}
\end{equation}

\noindent
where $F_{i,c}$ denotes the extracted value of color channel $c$ at pixel $i$ using stimuli snapshot ($S_t$) and head rotation ($R_t$) at time $t$, $P$ denotes the total number of pixels in each extracted frame. Notably, when calculating frame difference measures for foveal regions, the dimensions of the frame may differ across timestamps as parts of the user’ approximated foveal region may extend beyond one’s FOV of the stimuli. We addressed this possible dimension mismatch by only considering the intersection of the foveal region covered by the two frames. In practice, using the gaze tracking data, we also skipped calculations of pairs of frames where the participant closed their eyes for at least one considered timeframe.

We now derive a measure that quantifies frame differences caused solely by movement within the stimuli, while still accounting for the differences in individual viewpoints (i.e., $FD_{stim}$). In other words, this measure of frame differences is the counterfactual of active head movement, that is how much the frame would have changed if the user did not rotate their head at a given timestamp. To do this, we propose the following formulation, again taken between time $t-1$ and $t$

\begin{equation}
\begin{aligned}
FD_{stim}(t) = \frac{1}{3P} \sum_{c \in \{R,G,B\}} \sum_{i=1}^P & \left| F_{i,c}(S_t,R_{t-1}) \right. \\
& \left. - F_{i,c}(S_{t-1},R_{t-1}) \right|
\end{aligned}
\end{equation}
\noindent
The only prominent change is that the first term in the absolute difference operation is now of frame derived of the next timestamp (i.e., $S_t$) while assuming no head movement since the frame prior (i.e., $R_{t-1}$), thereby allowing us to construct the counterfactual as noted earlier. To derive out the final component of how much frame difference is attributable to head movement (i.e., $FD_{head}$), we define it as the difference between total frame difference and the amount attributable to the stimuli and is formalized as follows

\begin{equation}
FD_{\text{total}}(t) = FD_{\text{stim}}(t) + FD_{\text{head}}(t)
\end{equation}

\subsection{Categorizing virtual camera and narrative element movements}

We developed a coding scheme to label moment-to-moment dynamics of 360-degree videos, focusing on binary classification for movement in the virtual camera and narrative elements. While scholars have proposed solutions to automate similar processes \cite{jin2022}, we chose manual coding for three reasons. First, established frameworks for estimating virtual camera pose, such as Zhou et al. \cite{zhou2017}, are trained on 2D real-world videos, making it difficult to validate its legitimacy on our 360-degree videos with a large variation of artistic style (e.g., cartoon, abstract, realistic). Relatedly, because many automated approaches rely on vision-based techniques such as saliency maps \cite{huang2014, jin2022} and optical flow \cite{kale2015, zhang2018}, any association between the \emph{estimated} content characteristics and \emph{calculated} perceptual frame differences could merely reflect overlap in the signals used for deriving the two measures, rather than a meaningful relationship between them. Third, there are nuances to labeling movement in camera and narrative elements. For example, a pre-trained machine learning model may struggle to determine whether an object should be treated as a narrative element or background detail, or whether a view inside a moving car constitutes camera movement. As such, we deemed that a top-down rule-based manual labeling approach was better suited to our research goals.

The labeling process went as follows. Two authors independently went through all ten immersive videos, and performed initial labeling of the presence of camera and narrative elements in five second increments using the coding scheme introduced next. We refined the coding scheme once after both coders labeled one of the ten pieces. The initial code achieved substantial agreement for both camera movement ($\kappa_{camera}$=.91) and narrative element movement ($\kappa_{element}$=.73). Differences between the labels on the first piece were discussed until full consensus was achieved. Using the updated and final coding scheme, the two coders independently labeling the remaining nine pieces, achieving again substantial agreement prior to calibration ($\kappa_{camera}$=.89, $\kappa_{element}$=.69), and deliberated on their labels until full consensus was achieved. We present summary statistics for the four movement categories based on whether camera movement and narrative element movement were present or absent in a given five-second segment in Section \ref{sec: content char}.

\subsubsection{Virtual camera movement}
We quantified virtual camera movement as mismatches in perceived movement in a virtual environment and an individual’s vestibular responses can influence outcomes such as simulator sickness and disorientation in immersive contexts such as spatial exploration \cite{fernandes2016, rahimi2020}. We identified the presence of virtual camera movements as any five-second segments in which the viewpoint of the story changed, either through translation (i.e., x, y, z) or rotation (i.e., pitch, yaw, roll). Notably, this is labeled irrespective of the presence of visual anchors (e.g., the viewer’ relative position inside a car that does not change though the car is moving in the virtual environment). One exception to this is when visual anchors take up the entire frame such that it is impossible to determine from the perceptual information whether the viewing perspective is moving in the environment. For example, if the virtual camera is fixed in place inside a moving spaceship without any views into the outer space, these moments would be labeled as no virtual camera movement.

\subsubsection{Narrative element movement}
We also coded up in five-second increments whether narrative elements are moving on the screen, which echoes Jin et al.’s \cite{jin2022} focus on quantifying regions of interest using saliency maps derived from 360-degree videos. Specifically, we define narrative elements as visual elements that help to advance the narrative arc of the story, such as main characters or visual cues that guide the viewer’s attention to another area of the scene. Then, movement is marked by considerable changes in location of any of the narrative elements on the screen space in at least one of the x and y axes within the five second duration. For example, if the virtual camera is orbiting around a main character, keeping the character in the same place on the screen space, it would not be considered as movement; conversely, a character that traverses the screen space by moving around the virtual environment would be labeled as movement. 

\section{User study}
\subsection{360-Degree Videos}
To extrapolate generalizable insights on how viewers perceive 360-degree stories, we stimulus sampled ten 360-degree stories. Since the videos were streamed into a Unity program, such that physiological and behavioral data could be collected, we further narrowed the selection criteria of the stories to those that are accessible through public video streaming websites such as YouTube. From there, we selected 10 stories that were of relatively high resolution, longer than four minutes, and had narrative storylines. We chose video that is fictional to control for prior familiarity with video. Figure \ref{fig: stimuli} shows representative 360-degree screenshots from each piece, and the Appendix presents additional information regarding each stimulus.

\subsection{Participants}
We recruited 200 participants from a medium-sized private university using email, word-of-mouth, and department participant pool. Participants (Female=124, Male=71, Non-binary/third gender=2, Decline to answer=3) between ages 18 and 54 (M=23.36, SD=5.71) identified as Asian or Asian-American (N=114), Caucasian (N=36), Multi-racial (N=20), Hispanic or LatinX (N=16), African, African-American or Black (N=8), Middle Eastern (N=1), Indigenous/Native American, Alaska Native, First Nations (N=1), a racial group not listed (N=2), or declined to answer (N=2). They varied in prior VR experiences (None=33, One to two times=75, Three to five times=39, Six to ten times=14, More than ten times=39). The study was approved by the university’s IRB board. 

\subsection{Study setup and equipment}
\begin{figure}[t]
 \centering 
 \includegraphics[width=0.49\textwidth,keepaspectratio]{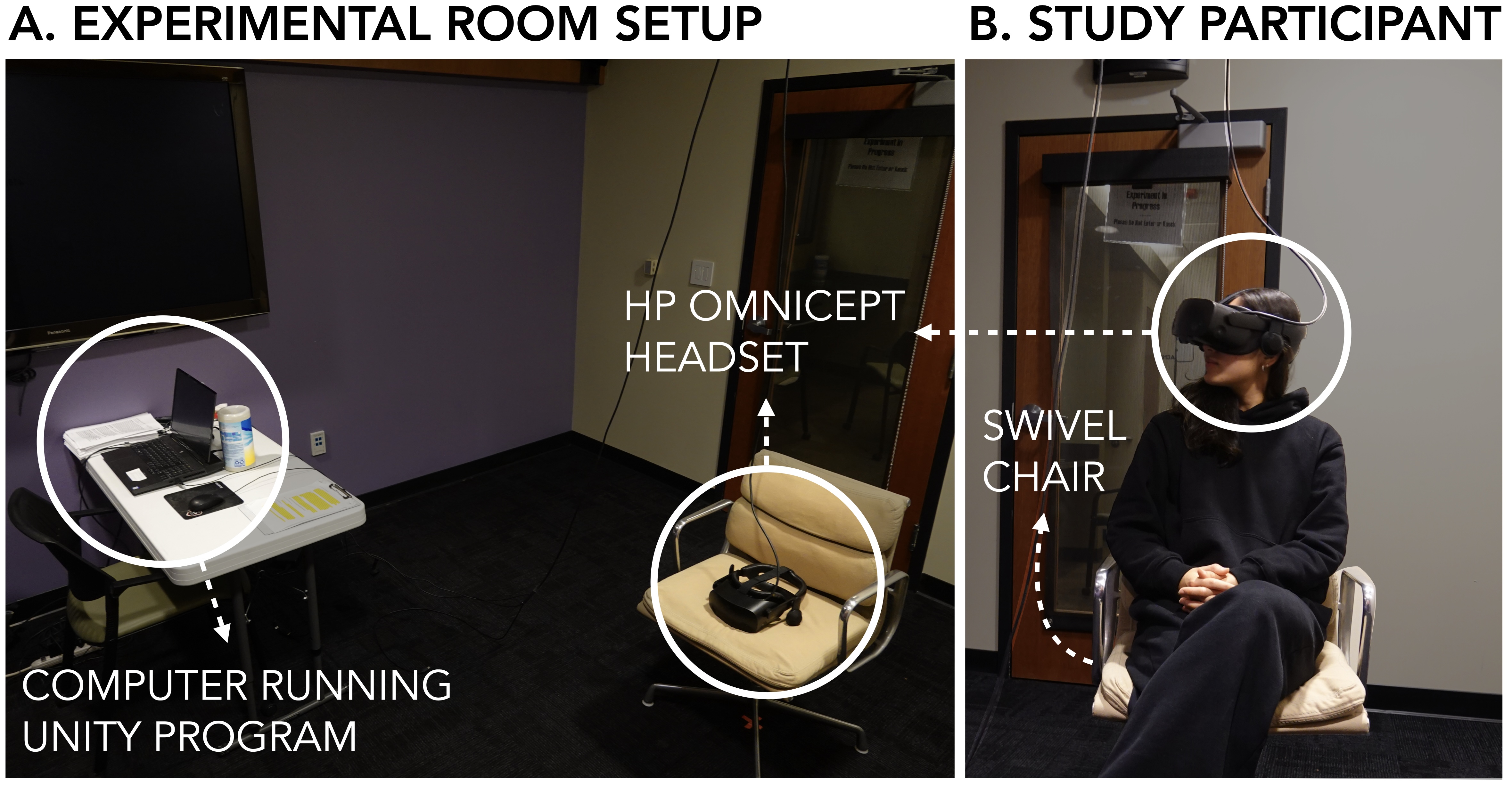}
 \caption{Overview of Study Setup. Panel A shows room setup and panel B shows an example of a user wearing the HP Omnicept headset. Areas of the images not highlighted by white circles are dimmed to improve visual clarity.}
 \label{fig: setup}
\end{figure}

Participants used the HP Omnicept Reverb G2 headset without handheld controllers. The headset has a diagonal field of view of approximately 114 degrees, and a per-eye resolution of 2160x2160. The headset uses a Tobii eye tracker and was powered by a laptop on an Intel CoreTM i9-9900K processor and NVIDIA GeForce RTX 2080 graphics card. During the VR component of the study, each participant was seated on a swivel chair that allowed them to rotate freely in the horizontal plane. The cord attaching the headset to the laptop is tethered through a ceiling hook to prevent tangling and tripping. Figure \ref{fig: setup} shows an example user in our study setup. We used two software programs. The first is the built-in HP eye and headset calibration program. The second is a program built using Unity 2023.2.20f1, which facilitated and collected physiological and behavioral data during the calibration and practice period, baseline measurements, and 360-degree video viewing. 

\subsection{Procedure}
\begin{figure}[t]
 \centering 
 \includegraphics[width=0.4\textwidth,keepaspectratio]{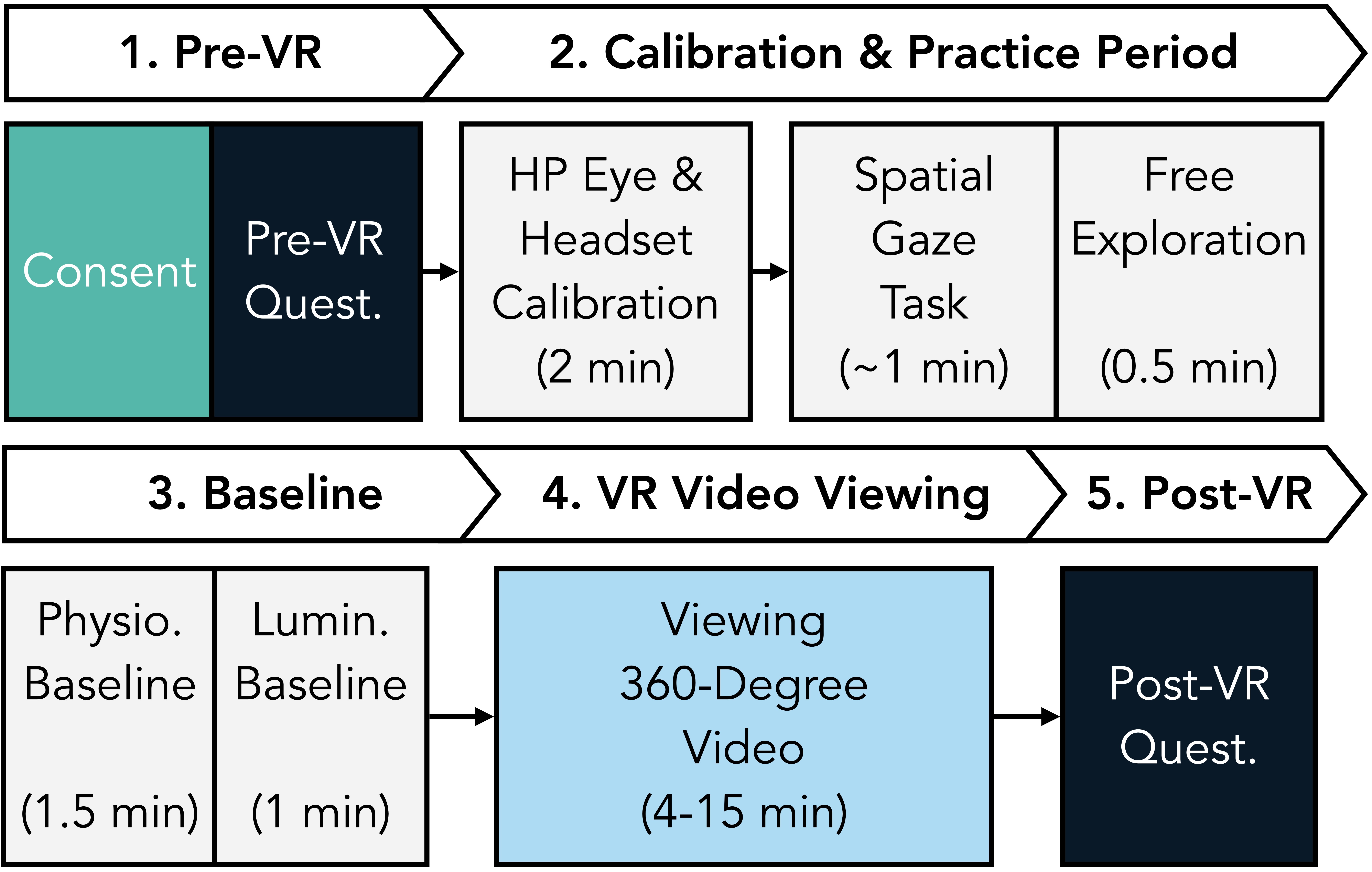}
 \caption{Overview of User Study Procedure.}
 \label{fig: procedure}
\end{figure}

The user study took 45--60 minutes, and consisted of five stages (Figure \ref{fig: procedure}). In the pre-VR component, participants completed the consent form and a pre-VR questionnaire on demographics and prior VR experiences. A researcher then checked with the participant on video genres they would prefer to avoid. If the participant was assigned a horror video and expressed discomfort with horror, we swapped out the video with a non-horror one. While this broke random assignment, it is better aligned with our goal of understanding how viewers naturally perceive immersive videos, as those who dislike horror are unlikely to engage with it outside a lab setting.

Then, a researcher fitted the VR headset onto the participant and began the HP eye and headset calibration program, which checked for the positioning of the headset on the participant’s head, adjusted for interpupillary distance, and performed an eye calibration. This process took around two minutes. 

Following this, the researcher ran the Unity program, which began with a practice period to familiarize the participant with moving in VR by looking around the virtual environment. For this, each participant completed a spatial gaze task, where they were instructed to look around the virtual environment to find a series of six red colored cubes. The red colored cubes were placed on each side of the square virtual environment, appeared one at a time, and turned green once the participant looked at them. Once all six cubes turned green, an additional thirty seconds were given to each participant to freely explore the environment. This procedure ensured that participants understood how to engage and look around virtual environments once they were immersed in the 360-degree video.

Participants then underwent two activities that served as baseline measures for physiological data. In the first activity, the participant was placed into the same virtual environment that they were in during calibration and was asked to remain still and not talk for 90 seconds. Then, to control for the effects of luminance on pupil dilation, we followed \cite{eckert2021} and \cite{chen2017}, and placed each participant into a space that went from complete darkness to increasingly brighter in five-second increments. The luminance of the space went up nine times, starting from 0 and stopping at 0.80 following \cite{burger2022}.

Following this, each participant viewed the 360-degree video. Before starting the video, the researcher reiterated the goal of the study and instructed the participant to watch the video as they normally would in a setting where they typically enjoy movies and videos, and that there are no tricks or hidden tasks. The participant was instructed to remain seated during the video but is allowed to rotate in the chair and look around the virtual environment as they wish. The researcher then started the video that lasted between 4 and 15 minutes.

After the video, participants filled out the post-VR questionnaire on topics such as emotions and narrative engagement.

\subsection{Self-report measures}
\subsubsection{Emotion} We measured valence and arousal using the self-assessment manikin (SAM) using a single-item 5-point scale (1=Lowest, 5=Highest), where participants are asked to select from a series of five images the image that most accurately represented them during the immersive experience \cite{bradley1994}. Mean ratings were 3.69 for valence ($SD_{valence}$=0.90) and 2.62 for arousal ($SD_{arousal}$=0.92). 

\subsubsection{Narrative engagement} Participants reported their narrative engagement levels using the 12-item 7-point narrative engagement scale \cite{busselle2009}, which we adapted for our study (e.g., swapping “program” with “360-degree story”). The scale has four subscales, each with three items. Items for each construct were averaged to form one measure per subscale, namely narrative understanding ($\alpha$=0.85, M=3.68, SD=1.59), attentional focus ($\alpha$=0.92, M=4.10, SD=1.63), narrative presence ($\alpha$=0.66, M=4.89, SD=1.14), and emotional engagement ($\alpha$=0.78, M=4.35, SD=1.37).

\subsubsection{Review}
Participants wrote a review of their immersive video experience. We followed marketing research literature reviewed in \cite{moon2021} and adopted it for our study. Three measures were collected. The first measure is an open-ended written review describing the experience. The second measure is a rating of the immersive experience on a single-item 7-point scale (1=Strongly Negative, 7=Strongly Positive; M=5.58, SD=1.08). The final measure assesses participants’ willingness to post their written review on a single 7-point scale (1=Strongly Unlikely, 7=Strongly Likely; M=3.58, SD=1.81).

\subsubsection{Cognitive load} We measured cognitive load using the NASA-TLX scale \cite{hart2006}, selecting four items that were most relevant. Notably, the original description for mental workload included both perceptual and mental demand. Given their different meanings in VR, we separated the item into two. The final items were mental demand, perceptual demand, physical demand, temporal demand, and frustration, with each item measured on a 7-point scale (1=Very Easy/Low, 7=Very Demanding/High). The five items are averaged to create one aggregate measure ($\alpha$=0.80, M=2.64, SD=1.12).

\subsubsection{Simulator sickness}Participants reported their levels of simulator sickness using the virtual reality sickness questionnaire \cite{kim2018}, which was evaluated using a 9-item 4-point Likert scale (0=Not at all, 3=Very). Following \cite{kim2018}, we aggregated the nine items into two scores for oculomotor and disorientation, and finally averaged the two to derive our measure for simulator sickness, which ranged from 0 to 100, with higher values indicating greater sickness (M=14.03, SD=14.92).

\subsection{Linguistic pattern of open-ended review}
We used Linguistic Inquiry and Word Count \cite{tausczik2010} to analyze open-ended reviews. Given our interests in the viewers’ subjective experiences, we extracted \emph{word count} (M=75.98, SD=39.58), as it reflects one’s investment and self-disclosure \cite{huang2022, wang2026}. We further extracted two affect measures, \emph{positive emotion} (M=1.66, SD=1.92), and \emph{negative emotion} (M=0.72, SD=1.29), which are derived by calculating the percentage of total words that match a given category (e.g., “happy” for positive emotion). Finally, we analyzed \emph{emotional tone}, which is a standardized variable on the scale of 1 to 100 that reflects the degree of positive vs. negative tone, with higher values indicating more positive tones (M=63.75, SD=31.40). 

\subsection{Behavioral and physiological measures}
With the HP Omnicept, we collected behavioral and physiological responses. As a result of the Unity program and the varying sampling rates of physiological sensors on the VR headset, head rotation and pupil dilation were logged at 30Hz, heart rate was recorded every 5 seconds, and heart rate variability was recorded every 60 seconds. Due to potential sensor drift and placement issues, the headset occasionally recorded invalid physiological data. Consequently, analysis with physiological measures was restricted to participants who maintained at least an 80\% data validity rate during both the baseline and video viewing periods. Table \ref{tab:physio validity} summarizes the validity of physiological data across participants.

\begin{table}[t]
    \centering
    \setlength{\tabcolsep}{2pt} 
    \small 
    
    \caption{\textsc{Summary Statistics of Physiological Data Validity. Columns show counts and percentages of participants with data validity falling under each bin of validity ranges.}}
    \label{tab:physio validity}
    
    \begin{tabular}{l >{\centering\arraybackslash}p{1.3cm} >{\centering\arraybackslash}p{1.2cm} >{\centering\arraybackslash}p{1.3cm} >{\centering\arraybackslash}p{1.6cm}}
        \toprule
        \textbf{Measure} & \textbf{[0, 0.4]} & \textbf{[0.4, 0.6]} & \textbf{[0.6, 0.8]} & \textbf{[0.8, 1]} \\
        \midrule
        Heart rate (video) & 3 (1.5\%)  & 1 (0.5\%) & 2 (1\%)   & 194 (97\%)   \\
        HRV (video)        & 16 (8\%)   & 2 (1\%)   & 12 (6\%)  & 170 (85\%)   \\
        Pupil Size (video) & 0 (0\%)    & 0 (0\%)   & 0 (0\%)   & 200 (100\%)  \\
        \addlinespace 
        Heart rate (baseline) & 10 (5\%)   & 5 (2.5\%) & 4 (2\%)   & 181 (90.5\%) \\
        HRV (baseline)        & 22 (11\%)  & 4 (2\%)   & 11 (5.5\%)& 163 (81.5\%) \\
        Pupil Size (baseline) & 1 (0.5\%)  & 1 (0.5\%) & 6 (3\%)   & 192 (96\%)   \\
        \bottomrule
    \end{tabular}
\end{table}

\subsubsection{Head rotation} From the head orientation (i.e., pitch, yaw, roll) collected during video viewing, we calculated the amount of head rotation. Head orientations were first converted into quaternions, and angular displacement between consecutive frames was then computed by deriving the relative quaternions between each pair of quaternions. We then summed up the magnitudes of the relative quaternions and divided them by the total video duration, yielding head rotation in degrees per second (M=22.01, SD=9.89).\footnote{We derived head movement by calculating the average displacement of the head in 3D space during video viewing. Head rotation and movement were highly correlated ($r$=0.95), so we chose to focus on head rotation to avoid multicollinearity during data analysis.}

\begin{table*}[t]
    \centering
    \small
    
    \caption{\textsc{Summary of Model Predicting Self-Report Measures and Linguistic Pattern. For brevity, we omitted three measures of narrative engagement (i.e., understanding, presence, emotional engagement). (*: $p \le .05$, **: $p \le .01$, ***: $p \le .001$).}}
    \label{tab:prediction results}
    \renewcommand{\arraystretch}{1.15}
    \setlength{\tabcolsep}{2pt}
    \begin{tabular}{l llllll llll}
        \toprule
        & \multicolumn{6}{c}{Self-Report Measure} & \multicolumn{4}{c}{Review Linguistic Pattern} \\
        \cmidrule(lr){2-7} \cmidrule(lr){8-11}
        Predictor & Valence & Arousal & Attentional & Cognitive & Simulator & Review & WC & Tone & Positive & Negative \\
        & & & Focus & Load & Sickness & Rating & & & Emotion & Emotion \\
        \midrule
        Head Rotation & $\uparrow$* & $\uparrow$ & $\uparrow$* & $\uparrow$ & $\downarrow$* & $\uparrow$ & $\uparrow$* & $\downarrow$ & $\downarrow$ & $\downarrow$ \\
        Heart Rate & $\downarrow$ & $\downarrow$ & $\uparrow$ & $\uparrow$ & $\uparrow$ & $\downarrow$ & $\downarrow$ & $\downarrow$ & $\downarrow$ & $\uparrow$ \\
        HRV & $\downarrow$ & $\uparrow$ & $\uparrow$ & $\uparrow$ & $\downarrow$ & $\downarrow$ & $\downarrow$ & $\uparrow$** & $\uparrow$ & $\downarrow$ \\
        Pupil Dilation & $\downarrow$* & $\uparrow$ & $\uparrow$ & $\downarrow$ & $\uparrow$ & $\downarrow$** & $\uparrow$ & $\downarrow$** & $\downarrow$ & $\uparrow$** \\
        \bottomrule
    \end{tabular}
    
    \vspace{5pt}
\end{table*}

\begin{table*}[t]
\centering
\small
\renewcommand{\arraystretch}{1.15}
\setlength{\tabcolsep}{5pt}

\caption{Effect Sizes and 95\% Confidence Intervals of Built Models. Models with significant findings are bolded.\\Values reported for review positive tone are of $R^2$ for the linear model.}
\label{tab:effect size}
\setlength{\tabcolsep}{6pt}
\begin{tabular}{llllll}
\toprule
Section & Model Outcome & $R^2_c$ & $CI_{c}$ & $R^2_m$ & $CI_{m}$ \\
\midrule

\ref{sec: prediction results} & \textbf{Valence} & 0.17 & [0.08, 0.34] & 0.09 & [0.03, 0.23] \\
    & Arousal & 0.03 & [0.03, 0.15] & 0.01 & [0.005, 0.10] \\
    & \textbf{Attentional Focus} & 0.28 & [0.12, 0.46] & 0.04 & [0.02, 0.14] \\
    & Emotional Engagement & 0.34 & [0.18, 0.52] & 0.03 & [0.01, 0.13] \\
    & Narrative Presence & 0.02 & [0.02, 0.14] & 0.01 & [0.01, 0.11] \\
    & Narrative Understanding & 0.52 & [0.28, 0.69] & 0.02 & [0.006, 0.09] \\
    & Cognitive Load & 0.03 & [0.02, 0.15] & 0.005 & [0.004, 0.09] \\
    & \textbf{Total Simulator Sickness} & 0.05 & [0.03, 0.19] & 0.04 & [0.02, 0.15] \\
    & \textbf{Review Rating} & 0.14 & [0.05, 0.31] & 0.07 & [0.02, 0.22] \\
    & \textbf{Review Word Count} & 0.10 & [0.04, 0.25] & 0.06 & [0.02, 0.19] \\
    & \textbf{Review Tone} & 0.14 & [0.08, 0.29] & 0.11 & [0.04, 0.22] \\
    & Review Positive Tone & 0.06 & [0.01, 0.12] & -- & -- \\
    & \textbf{Review Negative Tone} & 0.11 & [0.06, 0.26] & 0.09 & [0.04, 0.21] \\
\addlinespace
\ref{sec: visual scope} & \textbf{FD$_{\text{stim}}$} & 0.63 & [0.58, 0.68] & 0.37 & [0.32, 0.41] \\
\addlinespace
\ref{sec: content char} & \textbf{log(Head Rotation)} & 0.31 & [0.24, 0.41] & 0.03 & [0.03, 0.04] \\
    & \textbf{Heart Rate} & 0.74 & [0.69, 0.78] & 0.00 & [0.0001, 0.0008] \\
    & \textbf{Pupil Dilation} & 0.51 & [0.40, 0.63] & 0.03 & [0.02, 0.04] \\

\bottomrule
\end{tabular}
\end{table*}

\begin{figure*}[t]
 \centering 
 \includegraphics[width=0.88\textwidth,keepaspectratio]{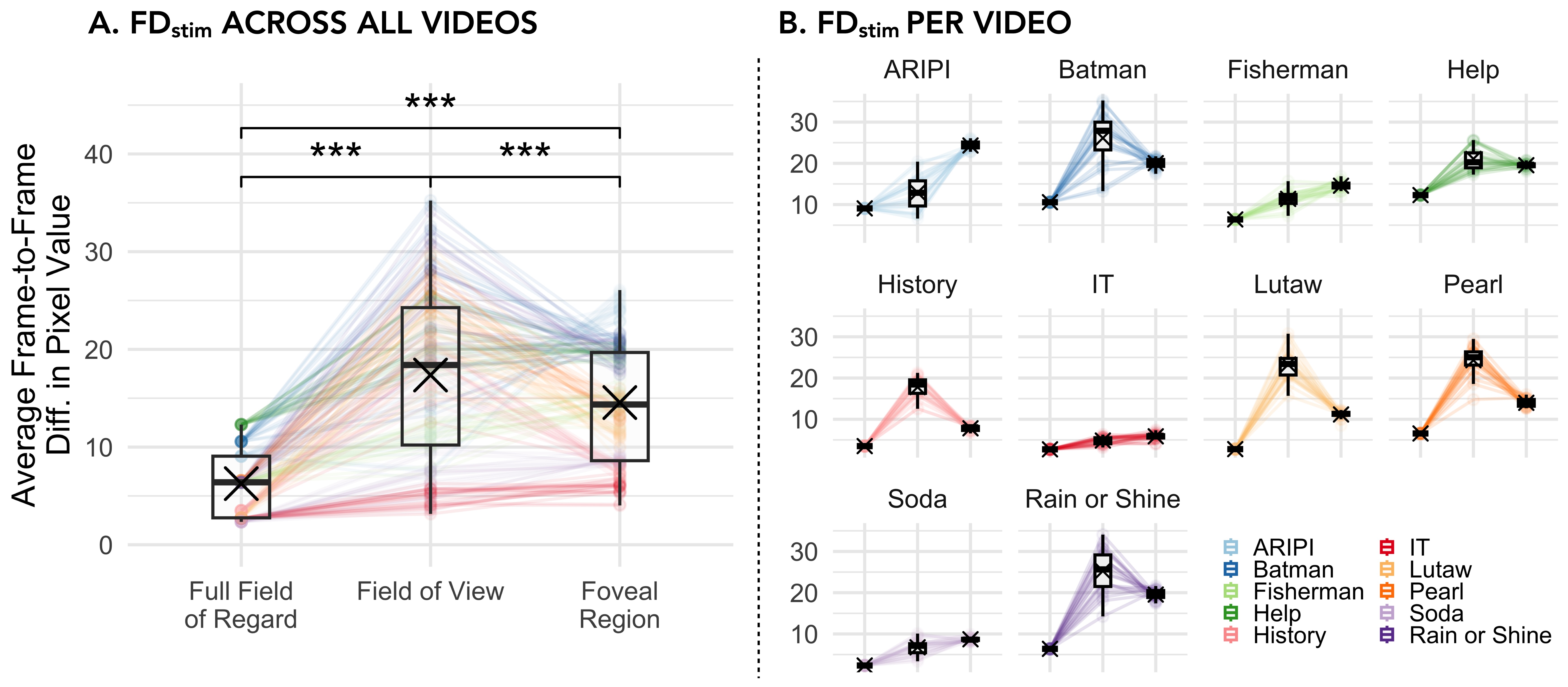}
 \caption{$FD_{stim}$ across Three Visual Scopes for All Ten Sampled Videos. (***: $p \le .001$)}
 \label{fig: frame difference results}
\end{figure*}

\subsubsection{Heart rate} We extracted the average heart rate, in beats per minute, during both the baseline and video viewing periods. Heart rate varies greatly across individuals, so we derived heart rate during video viewing as a percentage change from the average heart rate measured during the baseline.

\subsubsection{Heart rate variability (HRV)} The headset provided both the Root Mean Square of Standard Deviation between adjacent normal-to-normal (NN) intervals (RMSSD) and the Standard Deviation of NN intervals (SDNN). We focused on RMSSD for two reasons. First, it is a standard metric for assessing short-term stress and recovery \cite{zhao2023, kim2018hrv}. Further, as RMSSD and SDNN are highly correlated in our study (r=0.93 across baseline and video viewing), we selected only RMSSD to avoid multicollinearity during subsequent data analysis. Like heart rate, we baseline-adjusted the values for RMSSD by calculating the percentage change of RMSSD during video viewing compared to the baseline period to account for individual differences \cite{zhao2023}.

\subsubsection{Pupil dilation} Pupil dilation is reactive to the brightness of displays \cite{chen2017, eckert2021}, so we followed the approach in \cite{eckert2021} to correct for the effects of luminance on pupil sizes. For each participant, we first modeled the relationship between pupil dilation ($d$) and luminance ($L$). We calculated pupil dilation as the average of the left and right pupil diameters and modeled its relationship to luminance during the baseline period using

\begin{equation}
    d(L) = a \cdot d_{0} + \left( b \cdot \Delta d \frac{1}{1 + e^{(L - 0.25) \frac{c}{\Delta L}}} \right)
    \label{eq:pupil dilation 1}
\end{equation}

\noindent
where $d_0$ represents the average pupil dilation at $L$=0, $\Delta d$ represents the absolute difference between the largest and smallest pupil size observed, and $L$ denotes the range of luminance during baseline. Parameters $a$, $b$, and $c$ are fitted with bounds used in \cite{eckert2021}. Pupil dilation at time $t$ after accounting
for differences in luminance ($d_{CL}$) is derived as
\begin{equation}
    d_{CL} = d_m-d(Y(t-0.5))
\end{equation}
\noindent
where the first term ($d_m$) represents the measured uncorrected pupil dilation across both eyes at time $t$, and the second term represents the predicted pupil dilation attributable to luminance, adjusted for the pupil’s reaction time using the model established in Equation \ref{eq:pupil dilation 1}. Our analysis of pupil dilation is then of the corrected pupil dilation $d_{CL}$ for each participant.

\section{Results}
\subsection{Behavioral and physiological signals predict affective and cognitive outcomes} \label{sec: prediction results}
We built linear-mixed effects models using R’s lmerTest package and predicted user experience measures using head rotation, percentage change in heart rate from baseline, percentage change in HRV (i.e., RMSSD) from baseline, and pupil dilation. Video is modelled as a random effect. For models that resulted in a singular fit, meaning that the random effect of video does not explain a considerable amount of the variance in the outcome, we report results from a linear model instead. Significance is evaluated at $\alpha$=.05. Table \ref{tab:prediction results} summarizes our main findings. Effect sizes and confidence intervals of the models were computed with bootstrapping using the performance package and presented in Table \ref{tab:effect size}.

For predicting emotion, we found that valence is positively predicted by head rotation ($b$=0.02, $SE$=0.01, $p$=.023) and negatively predicted by pupil dilation ($b$=-0.56, $SE$=0.23, $p$=.016). No predictors significantly predicted arousal. For the four subconstructs of narrative engagement, we found that a higher level of head rotation is associated with greater attentional focus towards the narrative stimuli ($b$=0.03, $SE$=0.02, $p$=.041). No predictors significantly predicted the remaining three subconstructs, namely emotional engagement, narrative presence, and narrative understanding ($p$s>.05). For measures of effort and discomfort, the model predicting cognitive load revealed no significant main effects. The model predicting the amount of simulator sickness found that participants who rotated their heads less tended to report greater levels of simulator sickness ($b$=-0.35, $SE$=0.15, $p$=.026). No other predictor was significant in predicting simulator sickness.

In terms of the review of the narrative video, pupil dilation negatively predicted the rating of the review ($b$=-0.73, $SE$=0.27, $p$=.008), and no predictors significantly predicted the willingness to post reviews ($p$s>.05). In predicting linguistic patterns, participants who rotated their heads more wrote longer reviews ($b$=1.01, $SE$=0.40, $p$=.013), with those who rotated their heads more writing longer reviews. The percentage of changes in HRV from the baseline marginally and negatively predicted word count ($b$=-0.20, $SE$=0.10, $p$=.051). In predicting the tone of the written review, we found that HRV significantly and positively predicted it ($b$=0.21, $SE$=0.08, $p$=.008) and that pupil dilation was significant and negatively predicted tone ($b$=-21.97, $SE$=6.84, $p$=.003). There were no main effects when predicting the usage of words associated with positive emotion in the review, and pupil dilation positively predicted the usage of words associated with negative emotion ($b$=1.00, $SE$=0.30, $p$=.002).

\subsection{Frame difference differs across visual scope} \label{sec: visual scope}

We now explore how frame differences can reflect perceptual experiences and viewing strategies. We first compared $FD_{stim}$  across the three visual scopes of full FOR, FOV, and foveal region to understand what users choose to focus on in the scene. $FD_{stim}$ measures the amount of perceptual changes resulting from the viewing perspective in the absence of active head and gaze movements, making it a suitable metric for understanding what users choose to focus on in the 360-degree video. We built a linear mixed effects model predicting $FD_{stim}$ using visual scope type as a categorical variable and participant ID as a random effect. We also built a model with crossed random effects of participants and video IDs. This resulted in a singular fit, so the more parsimonious model with only participant ID as the random effect was used. Post-hoc pairwise comparisons are performed with Bonferroni correction. Figure \ref{fig: frame difference results} the $FD_{stim}$ values across visual scope types and Table \ref{tab:effect size} presents effect sizes of the built model.

Visual differences varied significantly across visual scope types ($p$<.001), and the post-hoc analysis revealed significant differences across all comparisons, namely FOR and FOV ($b$=11.11, $SE$=0.41, $p$<.001), FOR and foveal region ($b$=8.30, $SE$=0.41, $p$<.001), and FOV and foveal region ($b$=2.82, $SE$=0.41, $p$<.001). In other words, participants tended to fit in their FOV areas of the 360-degree scene with greater movement, though their eyes tend to focus on areas within their FOV that have comparably less perceptual movement. 

\subsection{Frame Differences is Related to Content Characteristics}
Seeing as the trends between visual scopes varied considerably across videos (Figure 6), we further explored whether $FD_{stim}$ across different visual scopes are related to content characteristics at the video level. To do this, we averaged the binary camera movement and narrative element movement labels across each video, which yielded four measures of frequencies of each possible combination of the camera and narrative element movement labels (e.g., narrative element movement with no camera movement). We then grouped participants’ frame differences in their FOV and foveal region by video, and averaged them to derive one measure per video. We also calculated the video-level differences between the frame differences in participants’ foveal region and FOV to study whether their gaze strategy of focusing on areas in their FOV with comparably less perceptual movement (as detailed in Section \ref{sec: visual scope}) is related to content characteristics. 

\begin{figure}[t]
 \centering 
 \includegraphics[width=0.4\textwidth,keepaspectratio]{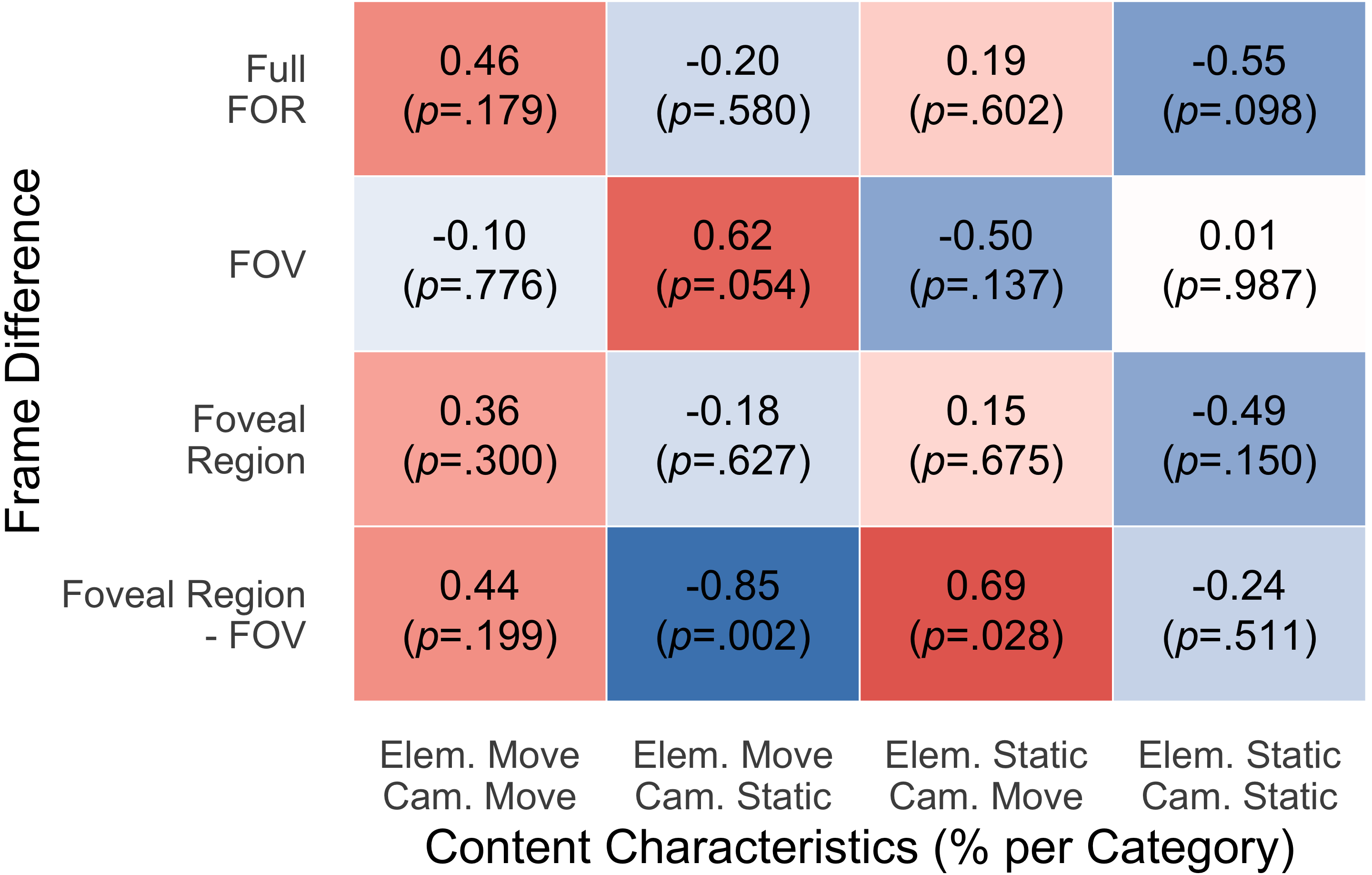}
 \caption{Spearman Correlations between Frame Differences and Content Characteristics. Results are based on ten data points, one per video. Content characteristics are labeled at the video level, and participant-level frame differences are averaged by video.}
 \label{fig: frame diff corr}
\end{figure}

Spearman correlations are performed between the four content characteristics categories and measures of frame differences. Frame differences across the entire FOR are also derived as a reference point. We report the main findings in Figure \ref{fig: frame diff corr}. Notably, participants who watched videos with greater percentages of segments with narrative element movement and no camera movement tended to adopt the strategy of looking at areas in their FOV with less perceptual movement ($\rho$=-0.85, $p$=.002). Conversely, for videos with more segments with camera movement and no narrative element movement, participants tended to focus on areas in their FOV with greater perceptual movement ($\rho$=0.69, $p$=.028). 

\subsection{Frame difference as a mediator for head rotation and user experience}

We examined whether the relationship uncovered between head rotation and user experiences can be explained by what users saw. For this, we performed mediation analyses testing whether the aggregated frame difference (i.e., $FD_{total}$) mediated the relationship between head rotation and the four outcome measures of valence, attentional focus, total simulator sickness, and review word count. We used the Mediation package in R with bootstrapping (3000 samples). We further tested whether $FD_{stim}$ and $FD_{head}$ mediated the relationships, and evaluated significance at $\alpha$=.05/2=.025.

$FD_{total}$ fully mediated the relationship between head rotation and valence (ACME=0.01, $p$<.001, 95\% CI[0.004, 0.02]; ADE=0.01, $p$=.160, 95\% CI[-0.004, 0.02]). The mediator accounted for 0.48 of the effect ($p$<.001, 95\% CI[0.17, 1.45]). $FD_{stim}$ also fully mediated this relationship (ACME=0.01, $p$<.001, 95\% CI[0.006, 0.02]; ADE=0.007, $p$=.400, 95\% CI[-0.009, 0.02]), with the mediator accounting for 0.66 of the total effect ($p$<.001, 95\% CI[0.26, 1.92]). $FD_{head}$ did not mediate this relationship (ACME=-0.006, $p$=.063, 95\% CI[-0.01, 0.00]; ADE=0.025, $p$<.001, 95\% CI[0.01, 0.04]; Prop. Mediated=-0.29, $p$=.063, 95\% CI[-1.03, 0.01]). 

Similar for simulator sickness, $FD_{total}$ fully mediated its relationship with head rotation (ACME=-0.08, $p$=.034, 95\% CI[-0.17, -0.01]; ADE=-0.23, $p$=.053, 95\% CI[-0.48, 0.00]), with the mediator accounting for 0.26 of the total effect ($p$=.036, 95\% CI=[0.01, 1.00]). $FD_{stim}$ fully mediated this relationship (ACME=-0.13, $p$=.007, 95\%, CI[-0.24, -0.04]; ADE=-0.18, $p$=.13, 95\%, CI[-0.42, 0.06]), contributing to 0.42 of the total effect ($p$=.009, 95\% CI[0.10, 1.46]), while the component of $FD_{head}$ did not (ACME=-0.02, $p$=.856, 95\% CI[-0.18, 0.09]; ADE=-0.29, $p$=.006, 95\% CI[-0.50, -0.09]; Prop. Mediated=0.06, $p$=.857, 95\% CI[-0.49, 0.53]).

Frame difference did not mediate the relationship between head rotation and the outcomes of attentional focus and review word count. $FD_{total}$ did not mediate the relationship between head rotation and attentional focus (ACME=0.002, $p$=.564, 95\% CI[-0.005, 0.01]; ADE=0.03, $p$=.016, 95\% CI[0.005, 0.05]; Prop. Mediated=0.07, $p$=.567, 95\% CI[-0.20, 0.59]). This was also true for the two subcomponents, $FD_{stim}$ (ACME=0.002, $p$=.691, 95\% CI[-0.009, 0.01]; ADE=0.03, $p$=.023, 95\% CI[0.003, 0.05]; Prop. Mediated=0.07, $p$=.693, 95\% CI[-0.35, 0.74]) and $FD_{head}$ (ACME=-0.005, $p$=.296, 95\% CI[-0.01, 0.00]; ADE=0.04, $p$=.004, 95\% CI[0.01, 0.06]; Prop. Mediated=-0.16, $p$=.301, 95\% CI[-0.74, 0.16]). 

$FD_{total}$ did not mediate the relationship between head rotation and review word count (ACME=0.08, $p$=.413, 95\% CI[-0.11, 0.27]; ADE=0.78, $p$=.009, 95\% CI[0.17, 1.38]; Prop. Mediated=0.09, $p$=.412, 95\% CI[-0.18, 0.40]). This is consistent with $FD_{stim}$ (ACME=0.18, $p$=.173, 95\% CI[-0.09, 0.44]; ADE=0.68, $p$=.027, 95\% CI[0.07, 1.29]; Prop. Mediated=0.21, $p$=.176, 95\% CI[-0.13, 0.74]) and $FD_{head}$ (ACME=0.21, $p$=.103, 95\% CI[-0.05, 0.47]; ADE=0.65, $p$=.071, 95\% CI[-0.06, 1.31]; Prop. Mediated=0.24, $p$=.114, 95\% CI[-0.07, 1.06]). 

\subsection{Content characteristics are associated with behavioral and physiological responses} \label{sec: content char}

Table \ref{tab:content characteristics} reports content characteristics across sampled videos. The distribution is well-balanced across categories derived from the binary labels for camera and narrative element movements. Using these labels, we analyzed whether content characteristics are related to behavioral and physiological responses. We built linear mixed-effects models predicting head rotation, heart rate, and pupil dilation. HRV was not predicted as HRV measured over ultra-short intervals is highly susceptible to measurement artifacts and is largely treated as a proxy for estimating longer-term indices rather than a direct moment-to-moment measure \cite{shaffer2020, fortrat1999, shaffer2017}. This puts to question whether HRV can reliably capture physiological reactions to five-second video segments. Methodologically, as HRV is sampled every 60 seconds, it introduces further complexities when aligning them to finer-grained labels of content characteristics and other behavioral and physiological signals. By contrast, heart rate provides a more reliable measurement compared to HRV in short time segments \cite{mcnames2006}.

\begin{table}[t]
    \centering
    \setlength{\tabcolsep}{4pt} 
    \small 
    
    \caption{\textsc{Summary Statistics of Content Characteristics.}}
    \label{tab:content characteristics}
    
    \begin{tabular}{l l >{\centering\arraybackslash}p{2.5cm}}
        \toprule
        Camera & Narrative Element & Count ($N$) \\
        Movement & Movement & and Percent (\%) \\
        \midrule
        No  & No  & 158 (20.31\%) \\
        No  & Yes & 231 (29.69\%) \\
        Yes & No  & 171 (21.98\%) \\
        Yes & Yes & 218 (28.02\%) \\
        \bottomrule
    \end{tabular}
\end{table}

Models were built in R using the lmerTest package. As camera and narrative element movements were annotated in five-second segments, we modeled moment-to-moment dynamics using binary indicators of narrative element and camera movements, along with time, as fixed effects. Time is operationalized as the starting time of the five-second segment in seconds, and is included to control for longitudinal effects of time. All models had at least one main effect for narrative element movement and camera movement, so we added an interaction term of narrative element movement and camera movement. To account for repeated measurements within participants and variability across videos, we specified crossed random effects for participant and video IDs. We winsorized data points where the predicted outcomes fall beyond three standard deviations of the sample mean and log-transformed head rotation as its raw values violated the normality assumptions of linear models. We present effect sizes of the three models in Table \ref{tab:effect size}.

For head rotation, there were positive main effects of narrative element movement ($b$=0.49, $SE$=0.02, $p$<.001) and camera movement ($b$=0.41, $SE$=0.03, $p$<.001), meaning that the presence of movement in both categories was associated with greater head rotation. Their interaction was significant ($b$=-0.34, $SE$=0.03, $p$<.001), in that the combined effect of narrative element and camera movement on head rotation was less than the sum of their individual effects. Head rotation decreased significantly over time ($b$=-0.0004, $SE$=0.00006, $p$<.001). For heart rate, there was a significant effect of camera movement ($b$=-0.53, $SE$=0.22, $p$=.017), with the presence of camera movement associated with decreases in heart rate. Narrative element movement was non-significant ($b$=0.13, $SE$=0.20, $p$=.52). Their interaction was non-significant ($b$=-0.05, $SE$=0.28, $p$=.868). Heart rate did not change significantly over time ($b$=0.0006, $SE$=0.0005, $p$=.229). Finally, for pupil dilation, narrative element movement was not significant ($b$=0.01, $SE$=0.01, $p$=.139) and camera movement was ($b$=-0.06, $SE$=0.01, $p$<.001), with the presence of camera movement associated with smaller pupil dilation. Their interaction was significant ($b$=0.11, $SE$=0.01, $p$<.001), meaning that the combined effect of narrative element and camera movement is greater than the sum of their independent effects on pupil dilation. Pupil dilation decreased significantly over time ($b$=-0.0005, $SE$=0.00002, $p$<.001).

\section{Discussion}
\subsection{Behavioral and physiological responses are related to subjective user experience}
Head rotation, HRV, and pupil dilation were meaningful predictors in predicting self-report user experiences and linguistic patterns. Most models showed small to medium effect sizes, with those for valence and review tone showing the largest effects. This provides evidence for the predictive value of these signals that are modest in practicality. Nevertheless, viewers who rotated their heads more reported greater valence, attentional focus, and wrote more in their subjective review. The finding on valence replicates prior work \cite{li2017}, and extends it by identifying perceptual change as a mechanism underlying this relationship. Specifically, our mediation analyses suggest two possible pathways linking head rotation to user experience. 

First, valence and simulator sickness were explained by a perceptual pathway: viewers who oriented toward more visually-dynamic parts of the scene experienced greater valence and less simulator sickness. This pattern holds for both total frame difference and the component attributable to the stimulus and user viewpoint, but not for the component driven by head movement itself. This suggests that these outcomes are shaped not by the active motion of moving one’s head, but by what that movement allows viewers to see. Second, head rotation predicted greater attentional focus and review word count after accounting for perceptual differences, reflecting a non-perceptual pathway. In this case, head rotation likely reflected exploratory engagement or top-down interest, as active looking is associated with greater involvement beyond the visual information acquired. It is, however, also possible that other perceptual processes explained these relationships, such as whether one’s gaze is tracking narrative elements.
Greater HRV was associated with more positively-toned written review, which suggests that viewers in more regulated autonomic and lower stress states were more likely to describe the experience positively \cite{brouwer2011, segerstrom2007}. Greater pupil dilation is associated with lower valence and video rating, as well as more negatively-toned reviews. To the best of our knowledge, our work is the first to demonstrate a significant relationship between pupil dilation and the numerical rating and tone of written reviews for immersive videos. While past works have shown that greater pupil dilation is linked to higher emotional arousal \cite{stanners1979, bradley2008}, as well as stronger consumer product preferences \cite{van2022} and willingness to pay \cite{ramsoy2017}, this positive association did not carry over to immersive video-watching experiences. Instead, greater pupil dilation likely signaled higher cognitive load \cite{eckert2021, lee2024, wei2025} or psychological states such as fear \cite{lee2023}, leading to greater pupil dilation being associated with more negative perceptions of the immersive experience.

\subsection{Content characteristics are related to perceptual differences and viewing behaviors}
Our coding of content characteristics revealed that narrative elements and camera movement are salient features of 360-degree videos, despite existing guidelines for immersive videos often noted the negative effects of moving the virtual camera \cite{meta2026, skarredghost2017}. Importantly, we found that these content features were related to a viewer’s perceptual experience, specifically whether their gaze tends to focus on areas of their field of view with greater perceptual differences (i.e., difference in $FD_{stim}$ between FOV and foveal region). Namely, viewers’ gaze were found to focus on areas with greater perceptual movement in their FOV when the video consisted of more moments of a moving camera accompanied by narrative elements that remained fixed in screen space. By contrast, when the video contained more moments of a fixed camera with movement in narrative elements, viewers’ gaze tended to focus on areas in their field of view with less perceptual movement. 

We interpret these findings as reflections of viewing strategies. In moments when the camera moved while narrative elements remained fixed, viewers’ gaze likely fell on perceptually-changing areas of the scene than on spatially-static narrative elements. This pattern could reflect a shift in viewer attention toward newly visible environmental information during such moments, as the stable position of narrative elements likely required less persistent visual attention. By contrast, when a camera is fixed and narrative elements moved, lower perceptual change in the foveal region relative to the FOV may indicate tracking of moving elements or fixation on more perceptually static parts of the scene. We posit that it is unnatural for viewers to focus on areas of high perceptual change in their FOV in these moments, as it would mean, for instance, focusing on part of the scene that a character is about to enter rather than the character itself. Consequently, focusing on any remaining parts of the scene will result in lower frame differences in one’s foveal region compared to their FOV.
 
Content characteristics predicted moment-to-moment head rotation, heart rate, and pupil dilation. Greater head rotation is associated with moments with camera and narrative element movement, though the negative interaction effect suggests a ceiling effect on how much head rotation can increase due to the presence of movement in both categories. The presence of camera movement reduced both heart rate and pupil dilation, which can signal lower cognitive load and arousal \cite{wascher2021, bradley1994}, though the effect size of heart rate was negligible in magnitude. One explanation is that when the camera moves for the viewer, it effectively guides attention through the scene, which in turn reduces the need for self-directed head movement and lowers cognitive effort and arousal. Another explanation is that the use of camera movement in 360-degree video more closely follows the conventions of traditional film, allowing viewers to draw on familiar heuristics and schemas developed for viewing 2D films, and in turn experience reduced cognitive load and arousal \cite{sweller2011}. The combination of camera movement and narrative element movement increased pupil dilation beyond the sum of their individual effects, suggesting that viewers experienced greater cognitive demands when both types of movement occurred concurrently.

\subsection{Practical Implications}
Behavioral and physiological signals can provide useful indicators of user experience in immersive media. Since VR systems capture metrics like head rotation, these signals may help developers assess engagement and discomfort and support iterative evaluation immersive experiences. Similar to how immersive systems can adapt to dynamically changing spatial and cognitive demands in everyday tasks \cite{lindlbauer2019, cheng2019}, immersive narrative content may also benefit from such adaptation. At the same time, our findings should be interpreted cautiously. The present analyses establish small-to-medium associations between behavioral and physiological responses and downstream subjective outcomes, but they do not demonstrate causality. To translate these patterns into more actionable guidelines, future research  should directly manipulate content features and presentation formats.

More broadly, the results underscore a central challenge in 360-degree content creation: no two viewers experience the same science in the same way. Our visualization of frame differences across and within content (Figure \ref{fig: frame difference results}) show considerable variability in what viewers attend to and how they follow scene dynamics. This variability in perceptual experiences is a defining feature of 360-degree media, but it is also consequential, as it relates to affective and cognitive outcomes such as valence and simulator sickness. For content intended to support open exploration, such variability may be desirable. However, for experiences that depend on directing attention toward particular events or characters, these findings point to the need to continue innovating on design strategies to guide viewer attention while preserving agency \cite{nielsen2016, rothe2019}.

This research offers key methodological insights. While prior works have shown that immersive content differs substantially in their broad characteristics \cite{paino2019, mabrook2019, de2022}, our findings suggest that it is also important to characterize how these features unfold from moment to moment. Our measure of frame differences provides one way to quantify these dynamics that account for perceptual changes attributed to the stimulus as well as the viewer’s head rotation. A similar point applies to content characteristics: understanding immersive experience requires not only identifying whether strategies such as camera or narrative movement are used for content, but also examining their prevalence, how they are applied over time, and how they interact with user behavior. From this perspective, content creators can likely benefit from modeling the relationship between content, perception, and viewer behaviors.

\section{Limitations and Future Works}

This work has several limitations. First, we relied on linear models to examine the directional relationships between behavioral and physiological signals and psychological outcomes. While this approach emphasizes interpretability, future work can apply deep learning methods to improve predictive accuracy \cite{wang2025, luong2022}. One may also model moment-to-moment behaviors, such as gaze and head movement trajectories, and benchmark performance on our dataset against existing approaches for attention modeling \cite{aladagli2017, xu2018, zhu2020}.

Our coding scheme for movement characteristics can be further developed. Future work can expand this framework to include categories such as visual anchors within moving scenes (e.g., sitting in moving car) and more fine-grained distinctions in camera motion such as different axes of rotation and translation as they have been previously shown to elicit different psychological outcomes \cite{keshavarz2011, bonato2009}. Related work could also benchmark machine learning approaches for automating these labels and study how more comprehensive content descriptors are related to behavioral, physiological, and self-report outcomes. Moreover, we quantified perceptual experience based on frame differences, which captures only one aspect of immersive videos. Scholars should explore alternative approaches for doing so, such as through areas of interest, visual saliency, and other measures of scene dynamics. Finally, the present work analyzed only a subset of the data we collected. We plan on extending these analyses to measures on memory recall and facial expressions.

\section{Conclusions}
This work analyzed how viewers watch 360-degree videos. We demonstrated that head rotation and physiological signals predicted cognitive and affective outcomes such as valence, simulator sickness, and tone of written reviews. We identified field of view perceptual changes as a mechanism shaping some of these relationships: those who rotate their heads more to experience greater frame-to-frame differences reported greater valence and lower simulator sickness. On viewing behaviors, individuals tended to orient their heads towards areas of the 360-degree scene with more perceptual changes, while their eyes focused on areas within their field of view with less perceptual changes. Results revealed that viewers adopted this strategy more in videos containing more moments with a static virtual camera accompanied by movement of narrative elements. We motivate research examining user experiences through moment-to-moment signals, and contribute to the foundations for creating meaningful immersive content.



\bibliographystyle{IEEEtran}
\bibliography{references}
\section{Biography Section}
\vspace{-33pt}

\begin{IEEEbiography}[{\includegraphics[width=1in,height=1.25in,clip,keepaspectratio]{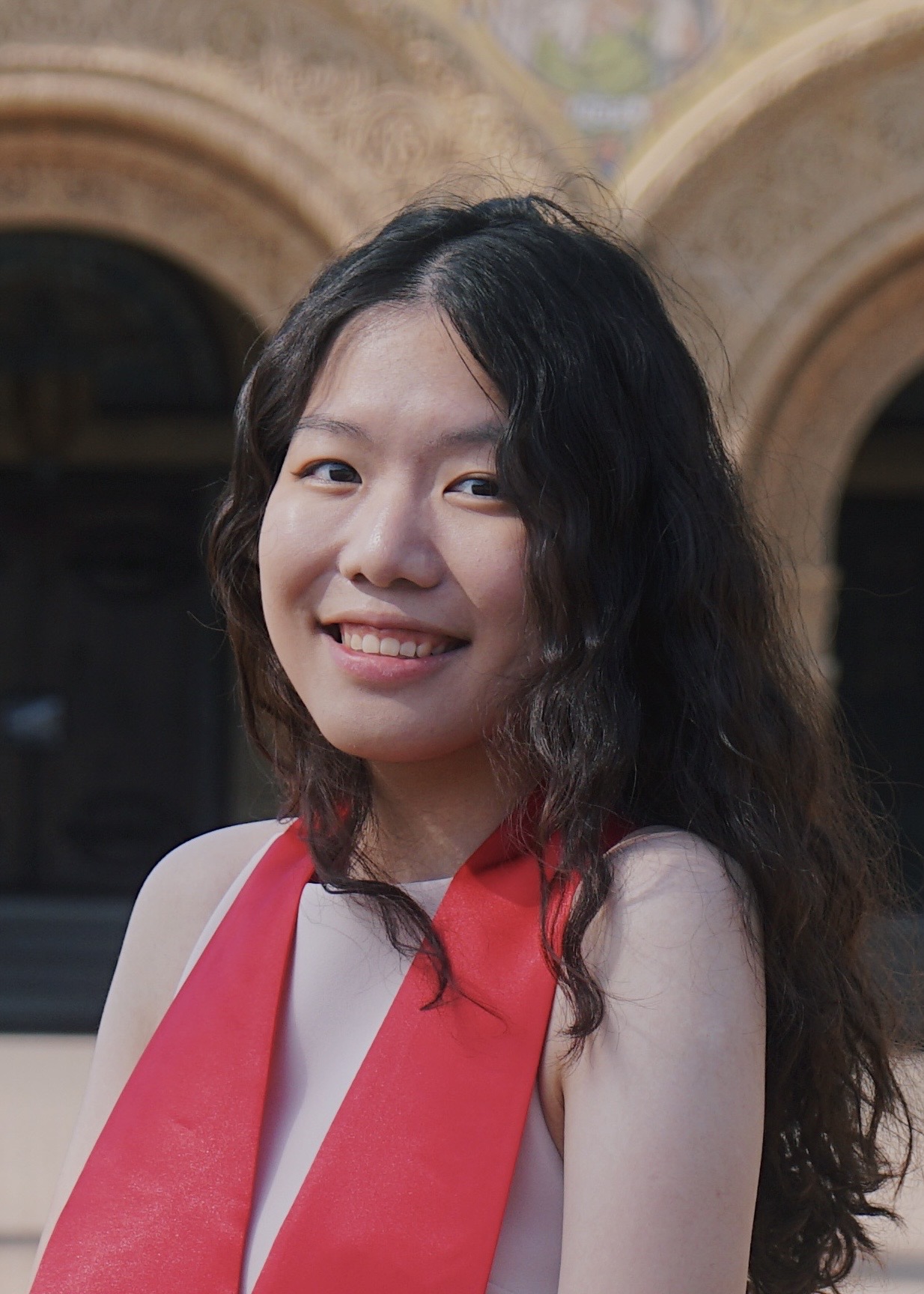}}]{Portia Wang} received the B.S. degree in Computer Science from Columbia University, New York, New York and her M.S. degree in Management Science and Engineering from Stanford University, Stanford, California. She is currently a Ph.D. Candidate in Communication at Stanford University, Stanford, California. Currently, she is investigating how individuals psychologically and behaviorally respond to immersive stories, and building tools that enable memory reliving using immersive technologies.
\end{IEEEbiography}
\vspace{-42pt}
\begin{IEEEbiography}[{\includegraphics[width=1in,height=1.25in,clip,keepaspectratio]{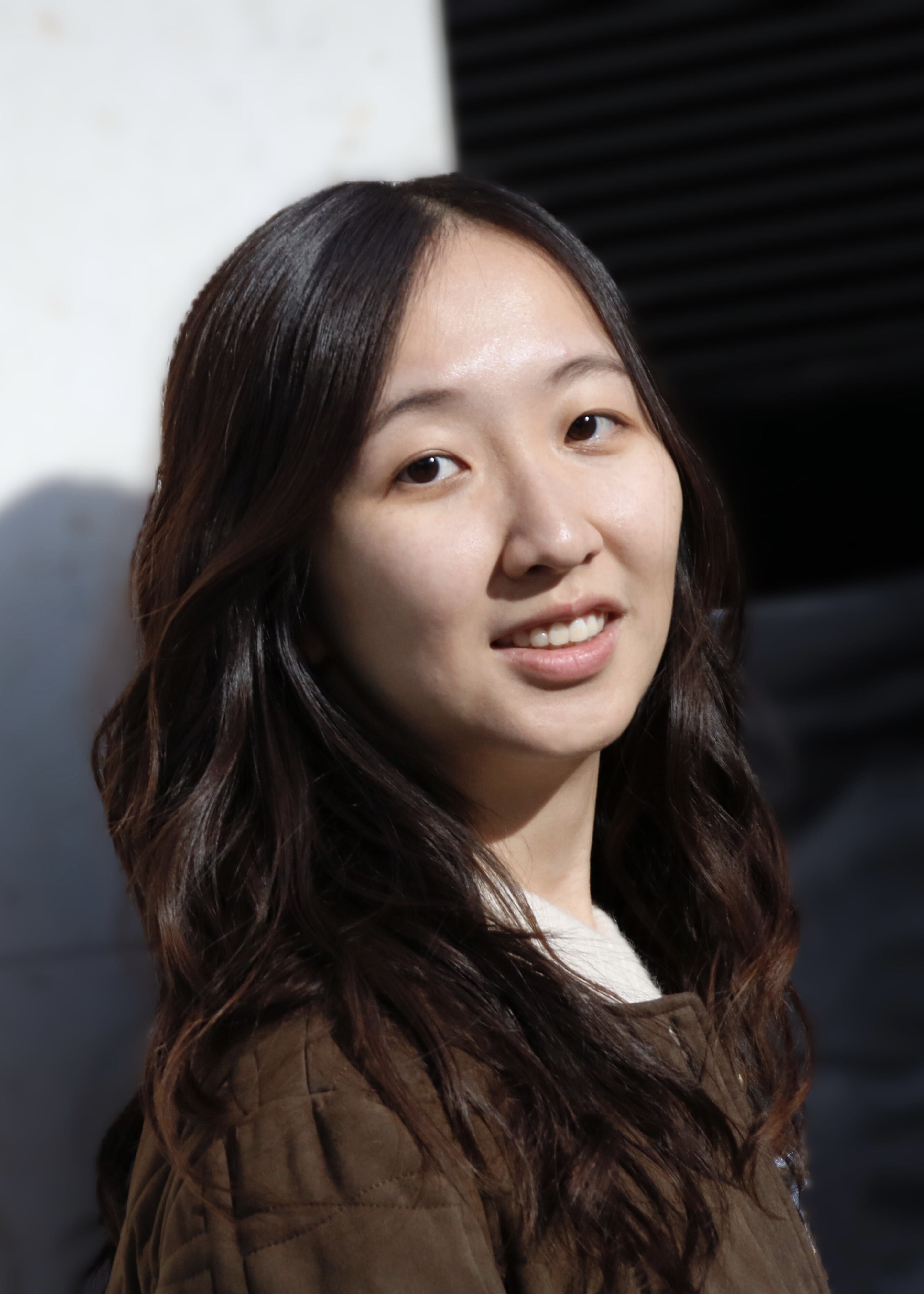}}]{Yujie Tao} recevied the B.A. in Journalism and Computer Science from the University of North Carolina at Chapel Hill, Chapel Hill, North Carolina and the M.S. degree in Computer Science from the University of Chicago, Chicago, Illinois. She is a Ph.D. Candidate in Computer Science at Stanford University, Stanford, California. My research focuses on multisensory wearable systems to support mental well-being and human performance.
\end{IEEEbiography}
\vspace{-35pt}
\begin{IEEEbiography}[{\includegraphics[width=1in,height=1.25in,clip,keepaspectratio]{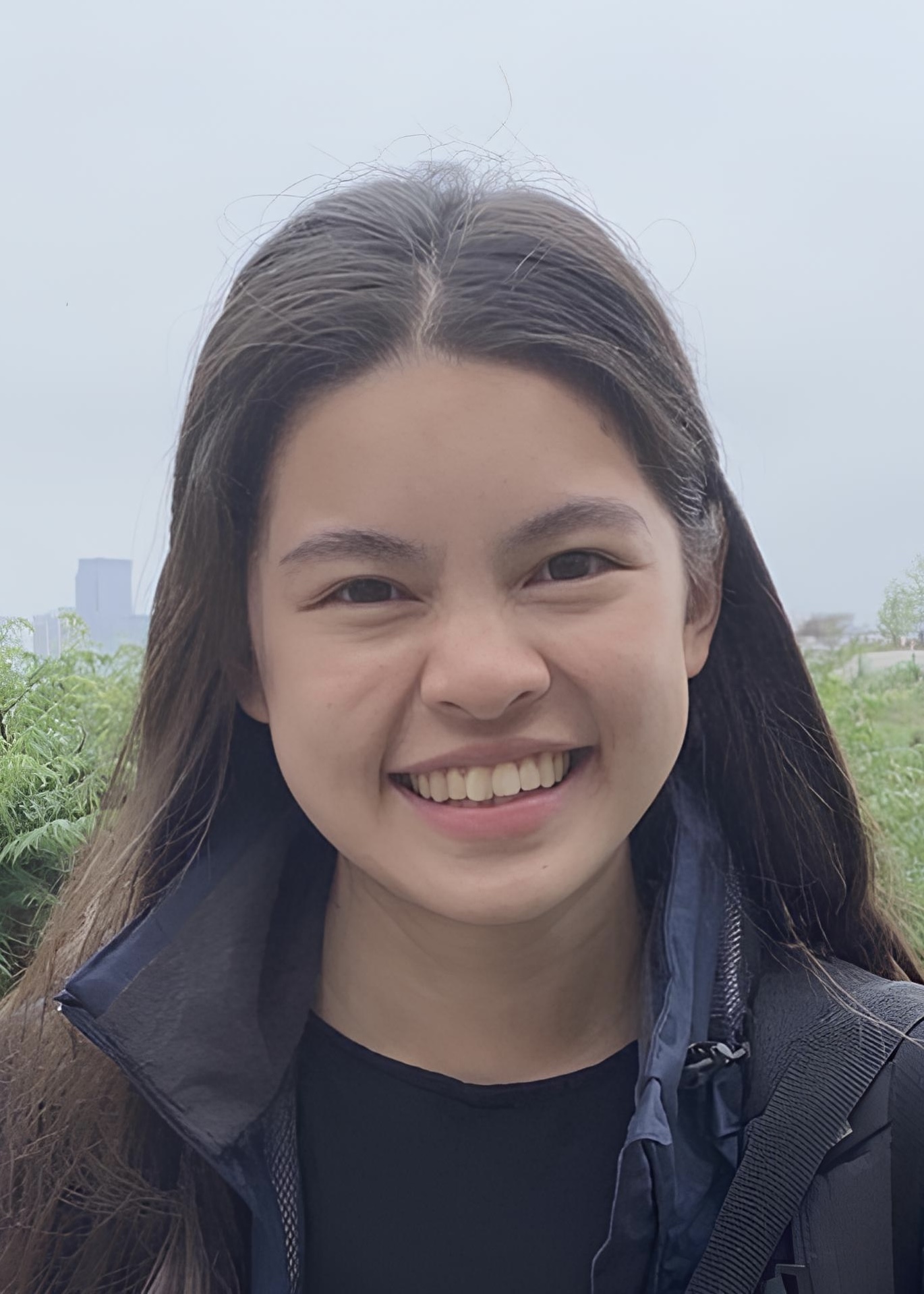}}]{Monique Santoso} received the B.A. degree in Environmental Science and Psychology from Middlebury College, Middlebury, Vermont and the M.A. degree in Communication from Stanford, Stanford, California. She is a Ph.D. Candidate in Communication at Stanford University. Her research focuses on the psychological, interpersonal, and behavioral implications of virtual reality, particularly as they relate to climate adaptation, prosocial attitudes and group interactions. 
\end{IEEEbiography}
\vspace{-42pt}
\begin{IEEEbiography}[{\includegraphics[width=1in,height=1.25in,clip,keepaspectratio]{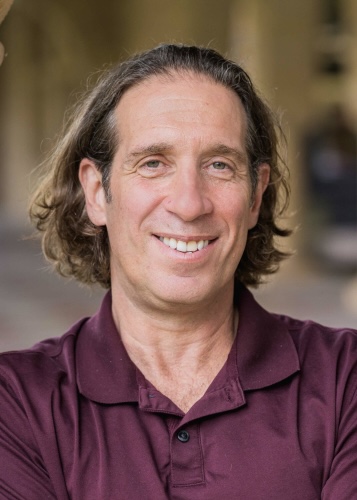}}]{Jeremy Bailenson} received the B.A. degree from the University of Michigan, Ann Arbor, Michigan, and the Ph.D. degree in cognitive psychology from Northwestern University, Evanston, Illinois. He is founding director of Stanford University’s Virtual Human Interaction Lab, Thomas More Storke professor with the Department of Communication, professor (by courtesy) of Education, professor (by courtesy) program in Symbolic Systems, a senior fellow at the Woods Institute for the environment, and a faculty leader at Stanford’s Center for Longevity. 
\end{IEEEbiography}

\vspace{11pt}

\vfill

\end{document}